\documentclass{article}

\usepackage{graphicx}

\usepackage{amsmath,latexsym}

\begin{document}

\setcounter{page}{1}

\pagestyle{plain}

\begin{center}
\Large{\bf Observational Viability of an Inflation Model with E-Model non-Minimal Derivative Coupling}\\
\small \vspace{1cm}  { Kourosh
	Nozari}$^{a,b,}$\footnote{knozari@umz.ac.ir}
\quad and \quad { Narges
	Rashidi}$^{a,b,}$\footnote{n.rashidi@umz.ac.ir}\\
\vspace{0.5cm} $^{a,b,}$Department of Physics, Faculty of Basic
Sciences,
University of Mazandaran,\\
P. O. Box 47416-95447, Babolsar, IRAN\\
$^{b}$ Research Institute for Astronomy and Astrophysics of Maragha (RIAAM),\\
P. O. Box 55134-441, Maragha, Iran\\
\end{center}

\begin{abstract}
By starting with a two-fields model in which the fields and their
derivatives are nonminimally coupled to gravity, and then by using a
conformal gauge, we obtain a model in which the derivatives of the
canonically normalized field are nonminimally coupled to gravity. By
adopting some appropriate functions, we study two cases with
constant and E-model nonminimal derivative coupling, while the
potential in both cases is chosen to be E-model one. We show that in
contrary to the single field $\alpha$-attractor model that there is
an attractor \textit{point} in the large $N$ and small $\alpha$
limits, in our setup and for both mentioned cases there is an
attractor \emph{line} in these limits that the $r-n_{s}$
trajectories tend to. By studying the linear and nonlinear
perturbations in this setup and comparing the numerical results with
Planck2015 observational data, we obtain some constraints on the
free parameter $\alpha$. We show that by considering the E-model
potential and coupling function, the model is observationally viable
for all values of $M$ (mass scale of the model). We use the
observational constraints on the tensor-to-scalar ratio and the
consistency relation to obtain some constraints on the sound speed
of the perturbations in this model. As a result, we show that in a
nonminimal derivative $\alpha$-attractor model, it is possible to
have small sound speed and therefore large non-Gaussianity.
\end{abstract}
\newpage

\section{Introduction}

Considering a single canonical scalar field (inflaton) responsible
for cosmological inflation in early universe, is a simplest way to solve some
problems of the standard model of cosmology. To have enough e-folds
number or equivalently enough exponential expansion of the
universe, the inflaton should rolls slowly down toward the minimum
of its potential. In this simple inflation model, an adiabatic,
scale invariant and gaussian mode of the primordial perturbations is
dominant~\cite{Gut81,Lin82,Alb82,Lin90,Lid00a,Lid97,Rio02,Lyt09,Mal03}.
However, it seems that in the future, with the advancement of
technology, we should be able to detect the non-Gaussian distributed
modes of the perturbations. Also, some extended models of inflation
predict scale dependent and non-Gaussian features of the primordial
perturbations~\cite{Mal03,Bar04,Che10,Fel11a,Fel11b,Noz12,Noz13a,Noz13b,Noz13c}.
In this regard, in studying the inflation, the models predicting the
non-Gaussian perturbations are really interesting.

It is possible the scalar field responsible for cosmological
inflation to be the Higgs boson. It is proposed that to adopt the
Higgs boson as the inflaton, we should consider a nonminimal
coupling between its derivatives and Einstein
tensor~\cite{Ame93,Ger10}. In this case, the friction is enhanced
gravitationally at higher energies because of the presence of
nonminimal derivative coupling. This means that the friction of an
inflaton rolling down its own potential increases significantly,
allowing occurrence of the slow-roll inflation even with steep
potentials. The models with nonminimal derivatives coupling are
capable to solve the unitary violation problem during inflation.
In these models, unitarity is not violated up to the quantum
	gravity scales and also, quantum gravity regime is avoided during
	Inflation~\cite{Ger10,Ger12a}. Note that, in~\cite{Ger10} it has
	been shown that to trust the effective inflationary description, the
	curvature should be much smaller than the Planck scale. Considering
	the relation $R=6(\dot{H}+2H^2)$ and the fact that during inflation
	era $H$ is nearly constant and $\dot{H}$ is very small, we have
	$R\simeq 12H^2$. So we can say that, to avoid the unitarity problem
	during inflation, $H$ should be much smaller than Planck mass and
	this is the unitarity bound. In the nonminimal derivative model this
	bound is not violated~\cite{Ger10,Ger12a,Ger12b}. Also, in these
models the perturbations are somewhat scale dependent and it is
possible to have non-Gaussian distributed perturbations. We refer
to~\cite{Tsu12,Sar10,Noz16a,Noz16b} for some works on the issue of
nonminimal derivatives in the early time accelerating expansion of
the universe as well as the late time cosmic dynamics.

Recently, the idea of ``cosmological attractor'' in the models
describing the cosmological inflation has attracted a lot of
attention. The conformal attractors~\cite{Kal13a,Kal13b} and
$\alpha$-attractors~\cite{Kai14,Fer13,Kal13c,Kal14} models are some
models which incorporates the idea of the cosmological attractors.
In
Refs.~\cite{Cec14,Kal13d,Kal14b,Lin15,Jos15a,Jos15b,Kal16,Sha16,Odi16,Ras18}
some more details on the issue of $\alpha$-attractors have been
studied. The conformal attractor model has the universal prediction
in the large e-folds number ($N$) for the scalar spectral index and
tensor-to-scalar ratio as $n_{s}=1-\frac{2}{N}$ and
$r=\frac{12}{N^{2}}$, respectively. The $\alpha$-attractor models
are divided into two categories named E-model and T-model, according
to the adopted potentials. The E-model corresponds to the following
potential
\begin{equation}
\label{eq1}
V=V_{0}\Big[1-\exp\big(-\sqrt{\frac{2\kappa^{2}}{3\alpha}}\phi\big)\Big]^{2n}\,,
\end{equation}
and the T-model is characterized by a potential as
\begin{equation}
\label{eq2}V=V_{0}\tanh^{2n}\Big(\frac{\kappa\phi}{\sqrt{6\alpha}}\Big)\,.
\end{equation}
In these potentials, $V_{0}$, $n$ and $\alpha$ are some free
parameters. The prediction of the scalar spectral index in the
$\alpha$-attractor model is similar to the prediction in the
conformal attractor ones as $n_{s}=1-\frac{2}{N}$ in small $\alpha$
and large $N$ limits. However, it predicts the tensor-to-scalar
ratio as $r=\frac{12\alpha}{N^{2}}$ in small $\alpha$ and large $N$
limits, which is somewhat different from the one predicted by the
conformal attractor model.

In this paper, we are going to study a nonminimal derivatives model
in which both the potential and nonminimal derivatives coupling
function are E-model type. Actually, the author of Ref.~\cite{Tsu12}
has studied the nonminimal derivatives model in which the coupling
function is a constant. He has adopted several types of potential
such as $\phi^{2}$, $\phi^4$, exponential and so on. Our attention
here is on $\phi^{2}$ potential. In Ref.~\cite{Tsu12}, it has been
shown also that the nonminimal derivatives model with $\phi^{2}$
potential is consistent with joint data of WMAP7~\cite{Kom11},
BAO~\cite{Per10}, and HST~\cite{Rie09} for $N=50$, $60$ and $70$.
However, if we compare the results with Planck2015 TT, TE, EE+lowP
data~\cite{Ade15a} the model is observationally viable just for some
values of $M$ (the mass scale of the nonminimal derivative
coupling). We are going to check that by considering E-type
potential and coupling function, whether the numerical results of
the model are consistent with Planck2015 TT, TE, EE+lowP data
background for all values of $M$ or not. In this regard, we follow
Refs.~\cite{Kal13a,Kal13b} and consider a model with two real scalar
fields $\psi$ and $\varphi$. The nonminimal action written in
Ref.~\cite{Kal13a} is conformal invariant. The authors in this
reference have used a $SO(1,1)$ conformal gauge (named rapidity
gauge) as
\begin{equation}\label{eq3}
\psi^{2}-\varphi^{2}=6
\end{equation}
which represents a hyperbola. By using a canonically normalized
field as
\begin{equation}\label{eq4}
\psi=\sqrt{6}\cosh (\frac{\phi}{\sqrt{6}}) \,\,, \quad
\varphi=\sqrt{6}\sinh (\frac{\phi}{\sqrt{6}})
\end{equation}
they were able to eliminate the nonminimal terms in the action and transform the
nonminimal action to the minimal one accordingly. Actually, by
adopting several types of the potential terms, they have obtained the models
corresponding to dS/AdS space, T-model of chaotic inflation and
Starobinsky model of inflation~\cite{Sta80,Sta83,Whi84}.

In our setup, both the fields and their derivatives are nonminimally
coupled to gravity. To eliminate the nonminimal coupling (
not the ``nonminimal derivatives coupling'') we use the gauge
(\ref{eq3}) and also rewrite equation (\ref{eq4}) as
\begin{equation}\label{eq5}
\psi=\sqrt{6}\cosh (\frac{\phi}{\sqrt{6\alpha}}) \,\,, \quad
\varphi=\sqrt{6}\sinh (\frac{\phi}{\sqrt{6\alpha}})
\end{equation}
where the free parameter $\alpha$ has been included which leads us
to E-model $\alpha$-attractor. Actually, $\alpha$ is
inversely proportional to the curvature of the inflaton K\"{a}hler
manifold~\cite{Kal13c}. By using this field, we
re-parameterize the two fields model and convert it to a one-filed
model with nonminimal derivatives coupling shown by the function
${\cal{F}}$. We study two cases as ${\cal{F}}=const.$ and
${\cal{F}}={\cal{F}}(\phi)$ and then we study cosmological inflation
and perturbations in this setup. We show that in both cases there is
an attractor \textit{line} in large $N$ and small $\alpha$ limits
which the $r-n_{s}$ trajectories tend to. Note that, as we said, in
the the single field $\alpha$-attractor model there is an attractor
\textit{point} in these limits. In Ref.~\cite{Noz17} it has been
shown that in the Gauss-Bonnet $\alpha$-attractor model also, there
is an attractor \textit{point} in the mentioned limits. Indeed, the
presence of line instead of point is because of considering the
nonminimal derivatives coupling which causes the scalar spectral
index of this model to be a functions of $\alpha$ and $M$. For
$M\rightarrow \infty$, we recover the attractor point in usual
$\alpha$-attractor models.

In section 2, we study the inflation in this NMDC $\alpha$-attractor
model and obtain the background equations of the model. In section 3,
by using the ADM formalism, we study the linear perturbations in this
model. In this section, we obtain some expressions for the scalar
spectral index and tensor-to-scalar ratio and compare the results
with Planck2015 observational data to test the observational
viability of the model. In this regard, we obtain some constraints
on parameter $\alpha$. In section 4, we study the non-linear
perturbations and non-Gaussian features of the primordial
perturbations. By using the relations between the tensor-to-scalar
ratio and sound speed, and using the allowed values of the
tensor-to-scalar ratio (from $95\%$ CL of Planck2015 TT, TE, EE+lowP
data), we obtain constraints on the sound speed in this model. These constraints
show that it is possible to have large non-Gaussianity in this
model. In section 5 we present a summary of our work.

\section{Inflation}

The action of a model with two real scalar fields $\psi$ and
$\varphi$, with both nonminimal and nonminimal derivatives couplings
to the gravity, is given by
\begin{eqnarray}\label{eq6}
S=\int
d^{4}x\sqrt{-g}\Bigg[\frac{\psi^2}{12\kappa^{2}}R+\frac{1}{2}\partial_{\mu}
\psi\,\partial^{\mu}\psi-\frac{\varphi^2}{12\kappa^{2}}R
-\frac{1}{2}\partial_{\mu}\varphi\,\partial^{\mu}\varphi+\frac{1}{2M^{2}}
\,f\,G^{\mu\nu}\partial_{\mu}\psi\,\partial^{\mu}\psi\nonumber\\-
\frac{1}{2M^{2}}\,F\,G^{\mu\nu}\partial_{\mu}\varphi\,\partial^{\mu}\varphi-U(\psi,\varphi)
\Bigg],
\end{eqnarray}
where the functions $f$ and $F$ show the nonminimal derivatives
coupling and are functions of the corresponding scalar field. Also, $U$ is a
potential term which is a function of the scalar fields. In the
absence of nonminimal derivatives coupling, the theory is
conformally invariant and by using the gauge $\psi^2-\varphi^2=6$
can be converted to the minimal coupling case~\cite{Fer13}. The
action (\ref{eq6}) by nonminimal derivatives coupling is a disformally invariant
action~\cite{Bet13}. Now, to proceed, we should specify all three
functions $f$, $F$ and $U$. We can adopt the nonminimal derivatives
coupling functions similar to the nonminimal coupling ones as
$f=\psi^{2}$ and $F=\varphi^2$. However, by this choice, if we use
gauge (\ref{eq3}) the nonminimal derivatives term vanishes (note
that, we use this gauge to eliminate the effect of nonminimal
coupling). We are not interested to this case because our aim in
this paper is to study the nonminimal derivatives model. So, we put
this case aside. On the other hand, since we seek for the effects of
the free parameter $\alpha$ on the observational viability of
nonminimal derivatives model by E-model potential and coupling
function, we should choose $f$ and $F$ in the way satisfying our
purpose. In another words, we should adopt appropriate
functions so that after using the gauge (\ref{eq3}) we reach a
nonminimal derivatives function with E-model type of the scalar
field's functions. To cover this purpose, it is convenient to
consider the same interacting function for $f$ and $F$ as
$f=F=\hat{{\cal{F}}}(\psi,\varphi)$. Now, by using the definitions
in equation (\ref{eq5}), we obtain
\begin{equation}\label{eq7}
\frac{\psi^{2}}{12}-\frac{\varphi^{2}}{12}=\frac{1}{2}\,,
\end{equation}
and
\begin{equation}\label{eq8}
\partial_{\mu}\psi\,\partial^{\mu}\psi-\partial_{\mu}\varphi\,\partial^{\mu}\varphi=
\partial_{\mu}\phi\,\partial^{\mu}\phi\,.
\end{equation}
We also define $\hat{{\cal{F}}}(\psi,\varphi)={\cal{F}}(\phi)$ and
$U(\psi,\varphi)=V(\phi)$. Now, we can rewrite the action
(\ref{eq6}) in the following form
\begin{eqnarray}
\label{eq9} S=\int
d^{4}x\sqrt{-g}\Bigg[\frac{1}{2\kappa^{2}}R+\frac{1}{2}\partial_{\mu}\phi\,\partial^{\mu}\phi
+\frac{1}{2M^{2}}{\cal{F}}(\phi)G^{\mu\nu}\partial_{\mu}\phi\,\partial^{\mu}\phi-V(\phi)
\Bigg],
\end{eqnarray}
In a spatially flat FRW metric, the action (\ref{eq9}) leads to the
following Friedmann equation
\begin{equation}
\label{eq10}H^{2}=\frac{\kappa^{2}}{3}\left[\frac{\dot{\phi}^{2}}{2}
\Big(1-\frac{9}{M^{2}}H^{2}{\cal{F}}\Big)+V(\phi)\right],
\end{equation}
where, a dot denotes a time derivative of the parameter. Variation
of the action (\ref{eq9}) with respect to the scalar field, gives
the equation of motion of the field $\phi$ as follows
\begin{eqnarray}
\label{eq11}
\ddot{\phi}\Big(-1+3\frac{{\cal{F}}H^{2}}{M^{2}}\Big)+\Big(\frac{6}
{M^{2}}{\cal{F}}H\dot{H}+\frac{9}{M^{2}}{\cal{F}}H^{3}-
3H\Big)\dot{\phi} +6{\cal{F}}'H^{2}\dot{\phi}^{2}-V'=0\,,
\end{eqnarray}
where, a prime refers to a field derivative of the parameter.
Considering the definitions $\epsilon=-\frac{\dot{H}}{H^{2}}$ and
$\eta=-\frac{1}{H}\frac{\ddot{H}}{\dot{H}}$, we obtain the slow-roll
parameters in our setup as
\begin{equation}
\label{eq12} \epsilon=\frac{Y}{1+\frac{\kappa^{2}}{2M^{2}}
	{\cal{F}}\dot{\phi}^{2}},
\end{equation}
\begin{eqnarray}
\label{eq13}
\eta=2\epsilon-\frac{\dot{Y}}{H\epsilon(1+\frac{\kappa^{2}}{2M^{2}}
	{\cal{F}}\dot{\phi}^{2})}
+\frac{Y}{H\epsilon}\frac{\kappa^{2}{\cal{F}}'\dot{\phi}^{3}+2\kappa^{2}{\cal{F}}\ddot{\phi}\dot{\phi}}
{2M^{2}(1+\frac{\kappa^{2}}{2M^{2}} {\cal{F}}\dot{\phi}^{2})^{2}}\,,
\end{eqnarray}
where parameter $Y$ is defined as
\begin{eqnarray}
\label{eq14} Y\equiv
\frac{\kappa^{2}{\cal{F}}\dot{\phi}\ddot{\phi}}{M^{2}H}+
\frac{V'^{2}}{2\kappa^{2}V^{2}}-3\frac{\kappa^{2}{\cal{F}}\dot{\phi}^{2}}{2M^{2}}
+\frac{\kappa^{2}{\cal{F}}'\dot{\phi}^{3}}{4M^{2}H}\,.\hspace{0.5cm}
\end{eqnarray}
To have inflation phase, the slow roll parameters should satisfy the
conditions $\epsilon \ll 1$ and $\eta \ll 1$. As soon as one of the
slow-roll parameter meets unity, the inflation ends. Satisfying the
conditions $\epsilon \ll 1$ and $\eta \ll 1$ means that the
conditions $\dot{\phi}^{2}\ll1$, $\ddot{\phi}\ll 3H\dot{\phi}$,
$\frac{9H^{2}|{\cal{F}}|}{2M^{2}}\dot{\phi}^{2}\ll V$ and
$|\dot{{\cal{F}}}|\ll |3{\cal{F}}H|$ should be satisfied.

In the inflation era, the Universe expands exponentially. This
expansion is characterized by the e-folds number defined as
\begin{equation}
\label{eq15} N=\int_{t_{hc}}^{t_{end}} H dt\,,
\end{equation}
where the subscripts $hc$ and $end$ denote the \emph{horizon crossing} and
\emph{end} of inflation respectively. In our setup, we find
\begin{equation}
\label{eq16} N\simeq \int_{\phi_{hc}}^{\phi_{end}} \frac{3H^{2}
	\Big(\frac{3{\cal{F}}H^{2}}{M^{2}}-1\Big)}{V'} d\phi\,.
\end{equation}

In the next section, we study the linear perturbations in the NMDC
setup to obtain the important perturbation parameters and compare to
the observational data.

\section{Linear Perturbations}

To study cosmological perturbations in this setup, we use the following ADM perturbed
line element
\begin{equation}
\label{eq17} ds^{2}=
-(1+2{\cal{R}})dt^{2}+2a(t){\cal{D}}_{i}\,dt\,dx^{i}
+a^{2}(t)\left[(1-2{\Phi})\delta_{ij}+2{\Theta}_{ij}\right]dx^{i}dx^{j}\,,
\end{equation}
where ${\cal{D}}^{i}=\delta^{ij}\partial_{j}{\cal{D}}+v^{i}$.
${\cal{R}}$ and ${\cal{D}}$ are 3-scalars. Also, the vector $v^{i}$
satisfies the condition $v^{i}_{,i}=0$~\cite{Muk92,Bau09}. The
spatial symmetric and traceless shear 3-tensor is denoted by
${\Theta}_{ij}$ and the spatial curvature perturbation is shown by
$\Phi$. Here, we consider the scalar part of the the perturbations
at the linear level and within the uniform-field gauge
($\delta\phi=0$), as
\begin{eqnarray}
\label{eq18}
ds^{2}=-(1+2{\cal{R}})dt^{2}+2a(t){\cal{D}}_{,i}\,dt\,dx^{i}
+a^{2}(t)(1-2{\Phi})\delta_{ij}dx^{i}dx^{j}\,,
\end{eqnarray}
to study the scalar perturbations. Using this perturbed metric and
expanding the action (\ref{eq9}) up to the second order in
perturbations gives the following quadratic action
\begin{equation}
\label{eq19} S_{2}=\int
dt\,d^{3}x\,a^{3}{\cal{W}}\left[\dot{\Phi}^{2}-\frac{c_{s}^{2}}{a^{2}}(\partial
{\Phi})^{2}\right],
\end{equation}
where
\begin{eqnarray}
\label{eq20}
{\cal{W}}=-4\frac{\left(\kappa^{-2}+\frac{\dot{\phi}^{2}{\cal{F}}}{2M^{2}}\right)^{2}\left(9\kappa^{-2}H^{2}
	-\frac{3}{2}\dot{\phi}^{2}
	+\frac{27H^{2}\dot{\phi}^{2}{\cal{F}}}{M^{2}}\right)}{3\left(
	2\kappa^{-2}H+\frac{3H\dot{\phi}^{2}{\cal{F}}}{M^{2}}\right)^{2}}
+3\left(\kappa^{-2}+\frac{\dot{\phi}^{2}{\cal{F}}}{2M^{2}}\right),
\end{eqnarray}
and
\begin{eqnarray}
\label{eq21}
c_{s}^{2}=3\Bigg[2\Big(2\kappa^{-2}H+\frac{3H\dot{\phi}^{2}{\cal{F}}}{M^{2}}\Big)\Big
(\kappa^{-2}+\frac{\dot{\phi}^{2}{\cal{F}}}{2M^{2}}\Big)H
-\Big(2\kappa^{-2}H+\frac{3H\dot{\phi}^{2}{\cal{F}}}{M^{2}}\Big)^{2}
\frac{\kappa^{-2}-\frac{\dot{\phi}^{2}{\cal{F}}}{2M^{2}}}{\kappa^{-2}+\frac{\dot{\phi}^{2}{\cal{F}}}{2M^{2}}}\nonumber\\
+4\Big(2\kappa^{-2}H+\frac{3H\dot{\phi}^{2}{\cal{F}}}{M^{2}}\Big)
\frac{d\Big(\kappa^{-2}+\frac{\dot{\phi}^{2}{\cal{F}}}{2M^{2}}\Big)}{dt}
-2\Big(\kappa^{-2}+\frac{\dot{\phi}^{2}{\cal{F}}'}{2M^{2}}\Big)\,
\frac{d(2\kappa^{-2}H+\frac{3H\dot{\phi}^{2}{\cal{F}}}{M^{2}})}{dt}\Bigg]\nonumber\\
\Bigg[
\Bigg(9\Big(2\kappa^{-2}H+\frac{3H\dot{\phi}^{2}{\cal{F}}}{M^{2}}\Big)^{2}
-4\Big(\kappa^{-2}+\frac{\dot{\phi}^{2}{\cal{F}}}{2M^{2}}\Big)\Big(9\kappa^{-2}H^{2}-\frac{3}{2}\dot{\phi}^{2}
+\frac{27H^{2}\dot{\phi}^{2}{\cal{F}}}{M^{2}}\Big)
\Bigg)\Bigg]^{-1}\,.
\end{eqnarray}
The parameter $c_{s}^{2}$ is the squared sound speed. We
	note that in the case of ${\cal{F}}=0$ we recover the standard
	second order action~\cite{Mal03,Che10,Hua13}. The two-point
correlation function, which is used to survey the power spectrum of
the perturbations, is defined as
\begin{equation}
\label{eq22} \langle
0|{\Phi}(0,\textbf{k}_{1}){\Phi}(0,\textbf{k}_{2})|0\rangle
=(2\pi)^{3}\delta^{(3)}(\textbf{k}_{1}+\textbf{k}_{2})\frac{2\pi^{2}}{k^{3}}{\cal{A}}_{s}\,.
\end{equation}
The parameter ${\cal{A}}_{s}$ in the right hand side of the above
equation is dubbed the power spectrum and is given by
\begin{equation}
\label{eq23}
{\cal{A}}_{s}=\frac{H^{2}}{8\pi^{2}{\cal{W}}c_{s}^{3}}\,.
\end{equation}

An interesting parameter in studying the scalar perturbations is the
scalar spectral index of the primordial perturbations. This
parameter specifies the scale dependence of the perturbation. The
scalar spectral index is defined as
\begin{equation}
\label{eq24} n_{s}-1=\frac{d \ln {\cal{A}}_{s}}{d \ln
	k}\Bigg|_{c_{s}k=aH}= -2\epsilon-\frac{1}{H}\frac{d \ln
	\epsilon}{dt} -\frac{1}{H}\frac{d \ln c_{s}}{dt}\,.\hspace{1cm}
\end{equation}

To study tensorial perturbations, by considering the tensor
part of the perturbed metric (\ref{eq17}) we write the 3-tensor
${\Theta}_{ij}$ as
\begin{equation}
\label{eq25}
{\Theta}_{ij}={\Theta}_{+}\vartheta_{ij}^{+}+{\Theta}_{\times}\vartheta_{ij}^{\times},
\end{equation}
with two polarization tensors $\vartheta_{ij}^{(+,\times)}$
satisfying the reality and normalization
conditions~\cite{Fel11a,Fel11b}. The tensor part of the
perturbed metric gives the following expression for the second order
action for the tensor mode
\begin{eqnarray}
\label{eq26} S_{T}=\int dt\, d^{3}x\, a^{3}
{\cal{W}}_{T}\left[\dot{\Theta}_{+}^{2}-\frac{c_{T}^{2}}{a^{2}}(\partial
{\Theta}_{+})^{2}+\dot{\Theta}_{\times}^{2}-\frac{c_{T}^{2}}{a^{2}}(\partial
{\Theta}_{\times})^{2}\right],
\end{eqnarray}
where the parameters ${\cal{W}}_{T}$ and $c_{T}^{2}$ are defined as
\begin{equation}
\label{eq27}
{\cal{W}}_{T}=\frac{1}{4\kappa^{2}}\left(1+\frac{\kappa^{2}{\cal{F}}\dot{\phi}^{2}}{2M^{2}}\right),
\end{equation}
\begin{equation}
\label{eq28}
c_{T}^{2}=1-\frac{\kappa^{2}{\cal{F}}\dot{\phi}^{2}}{2M^{2}}\,.
\end{equation}
By following the method used in the scalar part, we achieve the following
expression for the amplitude of the tensor perturbations
\begin{equation}
\label{eq29}
{\cal{A}}_{T}=\frac{H^{2}}{2\pi^{2}{\cal{W}}_{T}c_{T}^{3}}.
\end{equation}
By using equations (\ref{eq27})-(\ref{eq29}), we find the tensor
spectral index in this setup as
\begin{equation}
\label{eq30} n_{T}=\frac{d \ln {\cal{A}}_{T}}{d \ln k}=-2\epsilon\,.
\end{equation}
The ratio of the amplitudes of the tensor mode versus the scalar
mode, the tensor-to-scalar ratio, is given by
\begin{equation}
\label{eq31}
r=\frac{{\cal{A}}_{T}}{{\cal{A}}_{s}}\simeq16c_{s}\epsilon.
\end{equation}

Up to this point we have obtained some linear perturbation
parameters. Performing a numerical analysis on these parameters
gives some perspective on the cosmological viability of the model.
In this regard, we should specify the general functions ${\cal{F}}$
and $V$. Following~\cite{Kal13a,Kal13b}, we define $U(\psi,\varphi)$
as follows\footnote{Note that this choice of the potential breaks the $SO(1,1)$ isometry, see \cite{Kal13d,Her15} for more details.}
\begin{equation}
\label{eq32}
U(\psi,\varphi)=\frac{\lambda}{4}\varphi^{2}(\varphi-\psi)^{2}
\end{equation}
which by using equation (\ref{eq5}) leads to
\begin{equation}
\label{eq33}
V(\phi)=C\left[1-\exp\Bigg(-\sqrt{\frac{2\kappa^{2}}{3\alpha}}\,\phi\Bigg)\right]^{2}
\end{equation}
which is the E-model potential. As we have stated previously, $\alpha$ is
	inversely proportional to the curvature of the inflaton K\"{a}hler
	manifold. Given that we want to study two
cases as ${\cal{F}}=const.$ and ${\cal{F}}={\cal{F}}(\phi)$, we
should define two types of function for
$\hat{{\cal{F}}}(\psi,\varphi)$. The first one is as
\begin{equation}
\label{eq34}
\hat{{\cal{F}}}(\psi,\varphi)=\frac{\lambda}{4}(\varphi^{2}-\psi^{2})^{2}
\end{equation}
leading to
\begin{equation}
\label{eq35} {\cal{F}}=9\lambda
\end{equation}
This means that by this type of $\hat{{\cal{F}}}(\psi,\varphi)$ and
potential defined by (\ref{eq33}), we have an inflation model in
which the NMDC coupling is constant and potential is E-model.
Another function which we consider for
$\hat{{\cal{F}}}(\psi,\varphi)$ is defined as
\begin{equation}
\label{eq36}
\hat{{\cal{F}}}(\psi,\varphi)=\frac{\lambda}{4}\varphi^{2}(\varphi-\psi)^{2}
\end{equation}
leading to
\begin{equation}
\label{eq37}
{\cal{F}}(\phi)=C\left[1-\exp\Bigg(-\sqrt{\frac{2\kappa^{2}}{3\alpha}}\,\phi\Bigg)\right]^{2}\,.
\end{equation}
where
\begin{equation}
\label{eq38}
C=\frac{9\lambda}{4}
\end{equation}
By this function and potential (\ref{eq33}) we have an inflation
model in which both the NMDC function and potential are E-model.

\begin{figure}[]
\begin{center}
\includegraphics[scale=0.37]{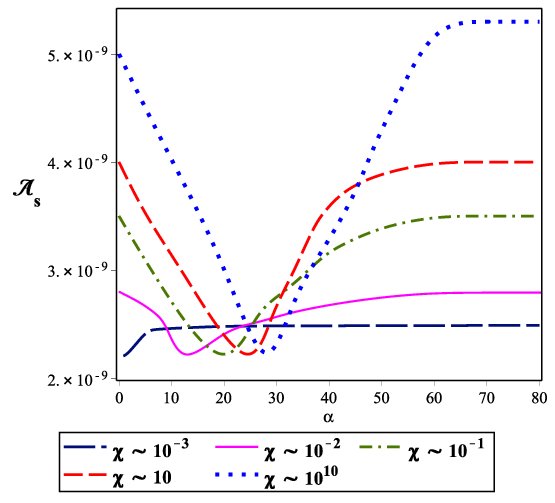}
\includegraphics[scale=0.37]{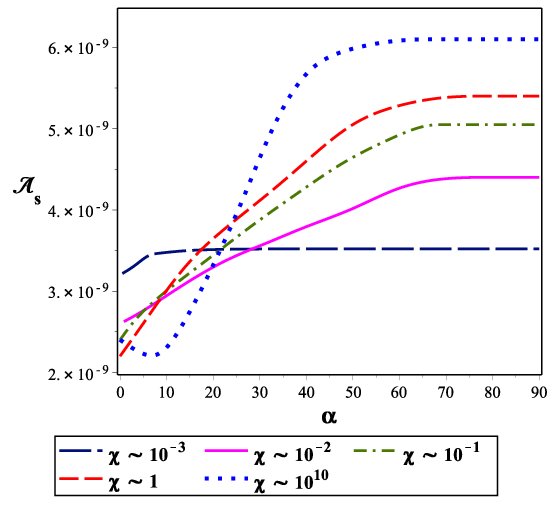}
\end{center}
\caption{\small {Power spectrum versus the free
parameter $\alpha$ in a nonminimal derivative model in which the
nonminimal coupling is a constant and the potential is E-model one. The
left panel corresponds to $N=60$ and the right panel is for $N=70$.}}
\label{fig1}
\end{figure}
\begin{figure}[]
\begin{center}
\includegraphics[scale=0.55]{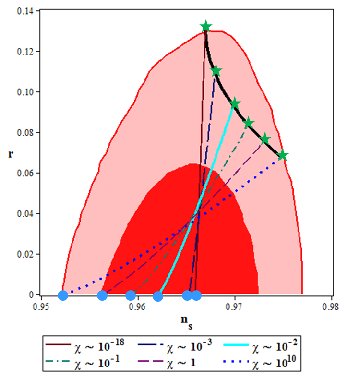}
\includegraphics[scale=0.56]{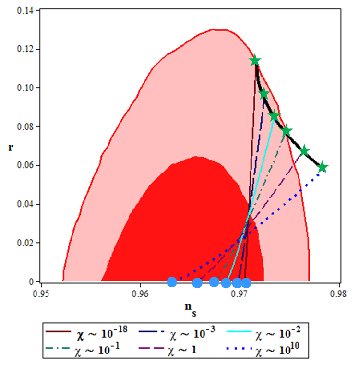}
\end{center}
\caption{\small {Tensor-to-scalar ratio versus the
scalar spectral index in a nonminimal derivative model in which the
nonminimal coupling is a constant and the potential is E-model one. The
left panel corresponds to $N=60$ and the right panel is
for $N=70$. The green stars show $\alpha\rightarrow
\infty$ limit and the blue circles show $\alpha\rightarrow 0$
limit.}}
\label{fig2}
\end{figure}

\begin{table*}
\begin{small}
\caption{\label{tab1}
Constraints on $\alpha$ by comparing the
scalar spectral index and tensor-to-scalar ratio with
Planck2015 TT, TE, EE+lowP data. These constraints are corresponding
to a nonminimal derivative model in which the nonminimal coupling is
a constant and the potential is E-model.}
\begin{tabular}{cccccccccc}
\\ \hline \hline&$\chi\sim 10^{-3}$&&$\chi\sim 10^{-2}$
&&$\chi\sim 10^{-1}$&&$\chi\sim 1$&&$\chi\sim 10^{18}$\\ \hline\\
$N=60\,,\,68\%$ CL& $\alpha< 58$ &&$\alpha< 61$&&$\alpha<
72$&&$\alpha< 76$ &&$\alpha<
85$\\\\
$N=60\,,\,95\%$ CL& all values of $\alpha$ &&all values of
$\alpha$&&all values of $\alpha$
&& all values of $\alpha$&&all values of $\alpha$\\\\
\hline\\ $N=70\,,\,68\%$ CL& $\alpha< 25$ &&$\alpha< 29$&&$\alpha<
37$
&&$\alpha< 44$&&$\alpha< 49$\\\\
$N=70\,,\,95\%$ CL&  all values of $\alpha$ && all values of
$\alpha$&& all values of $\alpha$
&&$\alpha< 102$&&$\alpha< 90$\\\\
\hline
\end{tabular}
\end{small}
\end{table*}

\begin{figure}[]
\begin{center}
\includegraphics[scale=0.37]{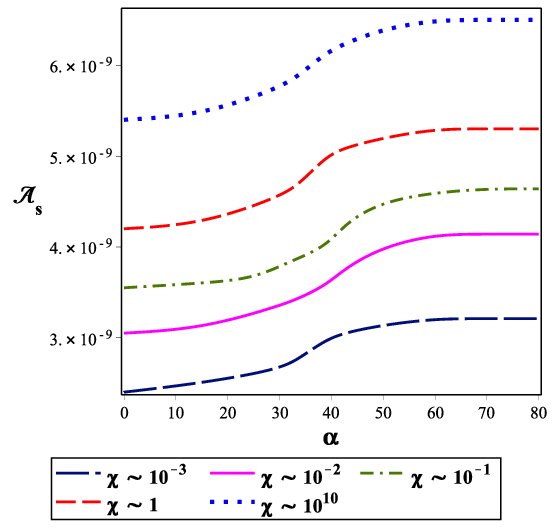}
\includegraphics[scale=0.37]{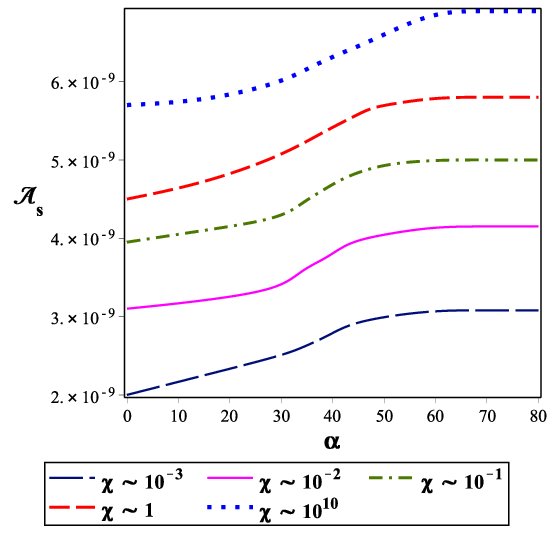}
\end{center}
\caption{\small {Power spectrum versus the free
parameter $\alpha$ in a nonminimal derivative model in which both
the nonminimal coupling and potential are E-model. The left panel
corresponds to $N=60$ and the right panel is for $N=70$.}}
\label{fig3}
\end{figure}

\begin{figure}[]
\begin{center}
\includegraphics[scale=0.55]{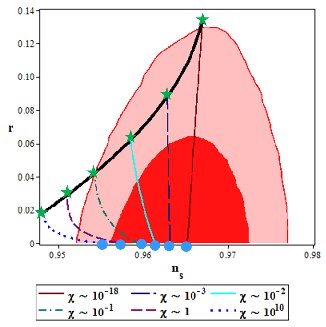}
\includegraphics[scale=0.56]{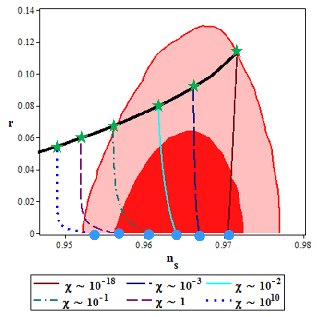}
\end{center}
\caption{\small {
Tensor-to-scalar ratio versus the
scalar spectral index in a nonminimal derivative model in which both
the nonminimal coupling and potential are E-model. The left panel
corresponds to $N=60$ and the right panel is for $N=70$.
The green stars show $\alpha\rightarrow \infty$ limit and
the blue circles show $\alpha\rightarrow 0$ limit.}}
\label{fig4}
\end{figure}

Given these equations, now we can study the model numerically for
both cases. By using $V$ and ${\cal{F}}$ defined as above, and
solving the integral (\ref{eq16}), we find the value of the field at
the Hubble crossing in terms of the model's parameters. By
substituting $\phi_{hc}$ in equations (\ref{eq23}), (\ref{eq24}) and
(\ref{eq31}), we perform our numerical analysis. We define
	parameter $\chi$ as $\chi\equiv\frac{C}{M^{2}}$ which is
	dimensionless since $\lambda$ (and therefore $C$ from Eq.
	(\ref{eq38})) has the dimension of mass squared. We adopt several
	sample values of $\chi$ in our numerical analysis. Note that, two
	limits of NMDC model are the GR limit (corresponding to
	$\chi\rightarrow 0$) and high-friction limit (corresponding to
	$\chi\rightarrow \infty$). We take $\chi\sim 10^{-3}$, $\chi\sim
	10^{-2}$, $\chi\sim 10^{-1}$, $\chi\sim 1$ and $\chi\sim 10^{10}$.
	The large value $\chi\sim 10^{10}$ is corresponding to the
	high-friction limit in NMDC model and it doesn't violate the
	unitarity bound as we have explained in Introduction. By these
	adoption, we study the evolution of the power spectrum versus
	$\alpha$ for $N=60$ and $N=70$. The results are shown in figure 1.
This figure corresponds to the functions defined in equations
(\ref{eq33}) and (\ref{eq35}). Since from Planck2015 data the power
spectrum is almost $2.4\times10^{-9}$, it seems that considering
E-model potential makes the power spectrum of the model to be
viable. For instance, depending on the value of parameter $\chi$,
the power spectrum takes the observed value with $10<\alpha<30$ for
$N=60$. For the case with $N=70$, we can get ${\cal{A}}_{s}\sim
2.4\times10^{-9}$ for $\chi <1$ and $\alpha<10$. However, in this
case also the power spectrum is of the order of $10^{-9}$ for the
adopted values of $\chi$. These situations are confirmed by
exploring $r-n_{s}$ map. Figure 2 shows the tensor-to-scalar ratio
versus the scalar spectral index for $N=60$ and $N=70$. Our analysis
shows that by considering the E-model potential, it is possible to
get $n_{s}=0.9652\pm 0.0047$ for all adopted values of $\chi$. For
$N=70$, the scalar spectral index is equal to $0.9652\pm 0.0047$
only with $\chi >1$. However, the presence of the free parameter
$\alpha$, causes $r-n_{s}$ plot with $N=70$ to lie in the background
of the Planck2015 TT, TE, EE+lowP data, for all values of $\chi$ and
in some ranges of $\alpha$. By comparing the values of the scalar
spectral index and tensor-to-scalar ratio with $68\%$ and $95\%$ CL
of Planck2015 TT, TE, EE+lowP data, we obtain some constraints on
parameter $\alpha$ which are summarized in table 1. Note that, as
figure 2 shows, in the NMDC model with E-model potential there is a
\emph{critical line} which the $r-n_{s}$ trajectories tend to. In
$M\rightarrow \infty$ limit, we recover the attractor points in
usual $\alpha$ attractor models (the line with $\chi\sim 10^{-18}$
shows almost this case), meaning that the presence of the line
instead of point is because of considering the nonminimal
derivatives coupling.

\begin{table*}
\begin{small}
\caption{\label{tab2} Constraints on $\alpha$ by comparing the
scalar spectral index and tensor-to-scalar ratio with
Planck2015 TT, TE, EE+lowP data. These constraints are corresponding
to a nonminimal derivative model in which both the nonminimal
coupling and potential are E-model.}
\begin{tabular}{cccccccccc}
\\ \hline \hline&$\chi\sim 10^{-3}$&&$\chi\sim 10^{-2}$
&&$\chi\sim 10^{-1}$&&$\chi\sim 1$&&$\chi\sim 10^{18}$\\ \hline\\
$N=60\,,\,68\%$ CL& $\alpha< 83$ &&$\alpha< 78$&&$\alpha<
55$&&$\alpha< 9$ &&all values of $\alpha$\\\\
$N=60\,,\,95\%$ CL& all values of $\alpha$ &&all values of
$\alpha$&&all values of $\alpha$
&& $\alpha< 46$ &&$\alpha<39$\\\\
\hline\\ $N=70\,,\,68\%$ CL& $\alpha< 85$ &&$\alpha< 80$&&$\alpha<
37$
&&$\alpha< 7$&&all values of $\alpha$\\\\
$N=70\,,\,95\%$ CL&  all values of $\alpha$ && all values of
$\alpha$&& all values of $\alpha$
&&$\alpha< 38$&&$\alpha< 11$\\\\
\hline
\end{tabular}
\end{small}
\end{table*}

We perform the same analysis for the NMDC model with functions
defined in (\ref{eq33}) and (\ref{eq37}). The results are shown in
figures 3 and 4. As these figures show, by adopting the E-model nonminimal
derivative coupling function, the NMDC model would be
observationally viable for all values of $\chi$ (even with large values such as
$\chi\sim 10^{10}$). Whereas, NMDC model with $\phi^{2}$ potential is
observationally viable just for some values of $\chi$. The power
spectrum in this model is of the order of $10^{-9}$ (for $\chi\sim
10^{-3}$ it is possible to get ${\cal{A}}_{s}\sim 2.4\times
10^{-9}$) and $r-n_{s}$ plane lies in the background of Planck2015
TT, TE, EE+lowP data (for some ranges of parameter $\alpha$). In this case
also, the $r-n_{s}$ trajectories tend to an \emph{attractor line} in
$\alpha\rightarrow 0$ limit. In the $M\rightarrow \infty$ limit we
recover the attractor points as usual in $\alpha$-attractor models. By
comparing the numerical results with $68\%$ and $95\%$ CL of
Planck2015 TT, TE, EE+lowP data, we obtain some constraints on the
free parameter $\alpha$ as shown in Table 2.

\section{Nonlinear Perturbations and Non-Gaussianity}

Regarding to the fact that by studying the two-point correlation
function we are not able to get information about non-Gaussian feature
of the primordial perturbations, then it is necessary to explore the
three-point correlation function. To this end, we expand the action
(\ref{eq9}) up to the third order in the small perturbations. By
introducing the new parameter ${\cal{X}}$ as follows
\begin{eqnarray}
\label{eq39}
{\cal{D}}=\frac{2(\kappa^{-2}+\frac{\dot{\phi}^{2}{\cal{F}}}{2M^{2}}){\Phi}}{2\kappa^{-2}H+3\frac{\dot{\phi}^{2}H{\cal{F}}}{M^{2}}
}+\frac{a^{2}{\cal{X}}}{\kappa^{-2}+\frac{\dot{\phi}^{2}{\cal{F}}}{2M^{2}}}\,,\hspace{1cm}
\end{eqnarray}
and
\begin{equation}
\label{eq40}
\partial^{2}{\cal{X}}={\cal{W}}\dot{\Phi}\,,
\end{equation}
we get the cubic action as
\begin{eqnarray}
\label{eq41} S_{3}=\int dt\, d^{3}x\,\Bigg\{
\Bigg[\frac{3a^{3}}{\kappa^{2}c_{s}^{2}}\,
\Bigg(1-\frac{1}{c_{s}^{2}}\Bigg) \epsilon
\Bigg]{\Phi}\dot{\Phi}^{2}
+\Bigg[\frac{a}{\kappa^{2}}\,\Bigg(\frac{1}{c_{s}^{2}}-1\Bigg)
\epsilon
\Bigg]{\Phi}\,(\partial{\Phi})^{2}\nonumber\\+\Bigg[\frac{a^{3}}{\kappa^{2}}\,\Bigg(\frac{1}{c_{s}^{2}\,H}\Bigg)
\Bigg(\frac{1}{c_{s}^{2}}-1\Bigg)\epsilon\Bigg]
\dot{\Phi}^{3}-\Bigg[a^{3}\,\frac{2}{c_{s}^{2}}\epsilon\dot{\Phi}
(\partial_{i}{\Phi})(\partial_{i}{\cal{X}})\Bigg]\Bigg\}\,,
\end{eqnarray}
up to the leading order in the slow-roll parameters. The three point
correlation function in the interaction picture is given by the
following expression
\begin{eqnarray}
\label{eq42} \langle
{\Phi}(\textbf{k}_{1})\,{\Phi}(\textbf{k}_{2})\,{\Phi}(\textbf{k}_{3})\rangle
=(2\pi)^{3}\delta^{3}(\textbf{k}_{1}+\textbf{k}_{2}+\textbf{k}_{3}){\cal{B}}_{\Phi}(\textbf{k}_{1},\textbf{k}_{2},\textbf{k}_{3})\,,
\end{eqnarray}
where
\begin{equation}
\label{eq43}
{\cal{B}}_{\Phi}(\textbf{k}_{1},\textbf{k}_{2},\textbf{k}_{3})=\frac{(2\pi)^{4}{\cal{A}}_{s}^{2}}{\prod_{i=1}^{3}
	k_{i}^{3}}\,
{\cal{G}}_{\Phi}(\textbf{k}_{1},\textbf{k}_{2},\textbf{k}_{3}).
\end{equation}
The parameter ${\cal{G}}_{\Phi}$ is given by
\begin{eqnarray}
\label{eq44}
{\cal{G}}_{\Phi}=\frac{3}{4}\Bigg(1-\frac{1}{c_{s}^{2}}\Bigg)
\Bigg(\frac{2\sum_{i>j}k_{i}^{2}\,k_{j}^{2}}{k_{1}+k_{2}+k_{3}}-\frac{\sum_{i\neq
		j}k_{i}^{2}\,k_{j}^{3}}{(k_{1}+k_{2}+k_{3})^{2}}\Bigg)
-\frac{1}{4}\Bigg(1-\frac{1}{c_{s}^{2}}\Bigg)\Bigg(
\frac{2\sum_{i>j}k_{i}^{2}\,k_{j}^{2}}{k_{1}+k_{2}+k_{3}}\nonumber\\-\frac{\sum_{i\neq
		j}k_{i}^{2}\,k_{j}^{3}}{(k_{1}+k_{2}+k_{3})^{2}}+\frac{1}{2}\sum_{i}k_{i}^{3}\Bigg)
+\frac{3}{2}\Bigg(\frac{1}{c_{s}^{2}}-1\Bigg)\Bigg(\frac{\left(k_{1}\,k_{2}\,k_{3}\right)^{2}}{(k_{1}+k_{2}+k_{3})^{3}}\Bigg).
\end{eqnarray}
One can use this parameter to define the so-called ``nonlinearity
parameter'' as
\begin{equation}
\label{eq45}
f_{NL}=\frac{10}{3}\frac{{\cal{G}}_{\Phi}}{\sum_{i=1}^{3}k_{i}^{3}}\,,
\end{equation}
which measures the amplitude of the non-Gaussianity of the
perturbations. By adopting different values of the three momenta
$k_{1}$ , $k_{2}$ and $k_{3}$, we can get different shapes of the
non-Gaussianity (see for instance~\cite{Wan00,Kom01,Bab04a,Sen10}).
In some inflation models (we can refer, for instance, the DBI,
k-inflation and higher derivative models) the non-Gaussianity is
constructed at horizon crossing during inflation epoch. In such
models, when all three wavelengths are equal to the size of the
horizon, there would be a maximal signal in the
bispectrum~\cite{Bab04b,Cre06}. In these models, it is useful to
study the ``equilateral" configuration of the non-Gaussianity. In
this setup and in the equilateral configuration we have

\begin{equation}
\label{eq46}
{\cal{G}}_{\Phi}^{equil}=\frac{17}{72}k^3\left(1-\frac{1}{c_{s}^{2}}\right)\,,
\end{equation}
leading to
\begin{equation}
\label{eq47}
f_{NL}^{equil}=\frac{85}{324}\left(1-\frac{1}{c_{s}^{2}}\right)\,.
\end{equation}

Now, we can analyze the model numerically. From equation
(\ref{eq47}) we see that the equilateral amplitude of the
non-Gaussianity corresponds to the sound speed. On the other
hand, the sound speed is related to the tensor-to-scalar ratio via
equation (\ref{eq31}). The evolution of the sound speed versus the
tensor-to-scalar ratio for models given by equations (\ref{eq35})
and (\ref{eq37}) are shown in figures 5 and 6 respectively. Given
that the tensor-to-scalar ratio is constrained by using the $95\%$
CL of Planck2015 TT, TE, EE+lowP data, we can find some constraints
on the sound speed in this model. The constrained are summarized in tables 3 and
4. As regard to the admissible values of $r$, it is possible to have
small sound speed. Since the amplitude of the equilateral
configuration of the non-Gaussianity is related to the sound speed
via equation (\ref{eq47}), with the small sound speed it is possible
to have large amplitude of the non-Gaussianity. In figures 7 and 8
we have plotted the equilateral configuration of the non-Gaussianity
versus the sound speed for some sample values of $\chi$ in the
background of Planck2015 TTT, EEE, TTE and EET data. Figure 7
corresponds to the function defined in equation (\ref{eq35}) and
figure 8 is corresponding to the one defined in equation
(\ref{eq37}). As we see, in some ranges of the model's parameter,
the amplitude of the non-Gaussianity in NMDC $\alpha$-attractor
model is consistent with Planck2015 TTT, EEE, TTE and EET data.

\begin{table*}
\caption{\label{tab3} Constraints on the sound speed and
tensor-to-scalar ratio obtained from $95\%$ CL of Planck2015 TT, TE,
EE+lowP data. These constraints are corresponding to a nonminimal
derivative model in which the nonminimal coupling is a constant and
the potential is E-model.}
\begin{tabular}{cccccccccc}
\\ \hline \hline&$\chi\sim 10^{-3}$&&$\chi\sim 10^{-2}$
&&$\chi\sim 10^{-1}$&&$\chi\sim 1$&&$\chi\sim 10^{18}$\\ \hline\\
$N=60\,\,$ & $r< 0.108$ &&$r< 0.092$&&$r<
0.083$&&$r< 0.074$ &&$r<0.066$\\\\
$N=60\,\,$ & $c_{s}^{2}<0.622$ && $c_{s}^{2}<0.527$&&
$c_{s}^{2}<0.457$
&& $c_{s}^{2}< 0.389$ &&$c_{s}^{2}<0.328$\\\\
\hline\\ $N=70\,\,$ & $r< 0.092$ &&$r< 0.082$&&$r<
0.075$&&$r< 0.06$ &&$r<0.046$\\\\
$N=70\,\,$ & $c_{s}^{2}<0.511$ && $c_{s}^{2}<0.448$&&
$c_{s}^{2}<0.391$
&& $c_{s}^{2}< 0.309$ &&$c_{s}^{2}<0.244$\\\\
\hline
\end{tabular}
\end{table*}

\begin{figure}[]
\begin{center}
\includegraphics[scale=0.35]{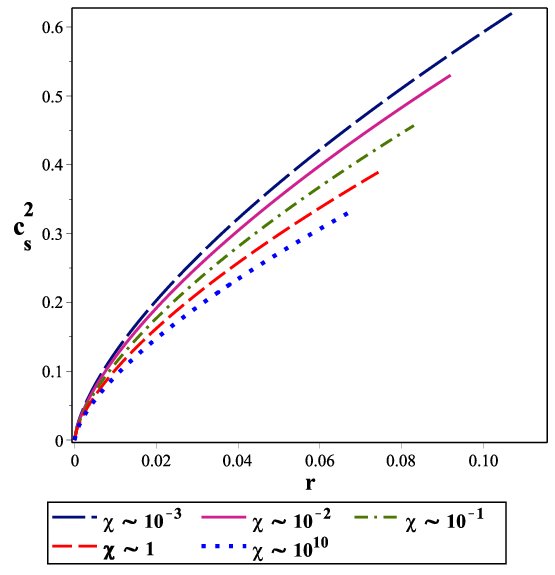}
\includegraphics[scale=0.35]{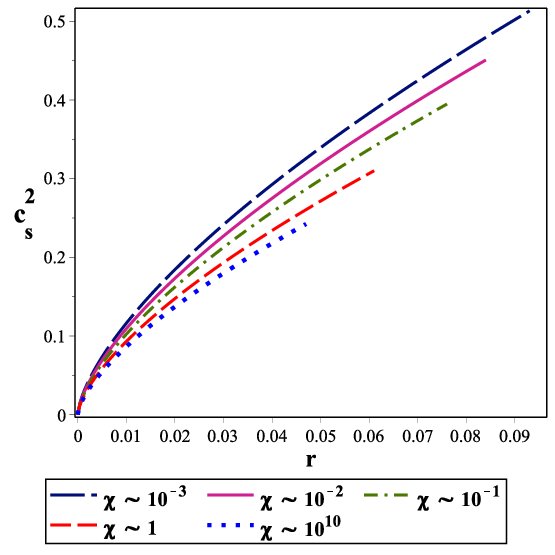}
\end{center}
\caption{\small {Sound speed versus the tensor-to-scalar
ratio in a nonminimal derivative model in which the nonminimal
coupling is a constant and the potential is E-model. The left panel
corresponds to $N=60$ and the right panel is for $N=70$.}}
\label{fig5}
\end{figure}

\begin{table*}
\caption{\label{tab4} Constraints on the sound speed and
tensor-to-scalar ratio obtained from $95\%$ CL of Planck2015 TT, TE,
EE+lowP data. These constraints are corresponding to a nonminimal
derivative model in which both the nonminimal coupling and potential
are E-model.}
\begin{tabular}{cccccccccc}
\\ \hline \hline&$\chi\sim 10^{-3}$&&$\chi\sim 10^{-2}$
&&$\chi\sim 10^{-1}$&&$\chi\sim 1$&&$\chi\sim 10^{18}$\\ \hline\\
$N=60\,\,$ & $r< 0.088$ &&$r< 0.061$&&$r<
0.039$&&$r< 0.029$ &&$r<0.018$\\\\
$N=60\,\,$ & $c_{s}^{2}<0.422$ && $c_{s}^{2}<0.312$&&
$c_{s}^{2}<0.213$
&& $c_{s}^{2}< 0.16$ &&$c_{s}^{2}<0.105$\\\\
\hline\\ $N=70\,\,$ & $r< 0.091$ &&$r< 0.080$&&$r<
0.065$&&$r< 0.059$ &&$r<0.055$\\\\
$N=70\,\,$ & $c_{s}^{2}<0.514$ && $c_{s}^{2}<0.444$&&
$c_{s}^{2}<0.365$
&& $c_{s}^{2}< 0.324$ &&$c_{s}^{2}<0.292$\\\\
\hline
\end{tabular}
\end{table*}

\begin{figure}[]
\begin{center}
\includegraphics[scale=0.35]{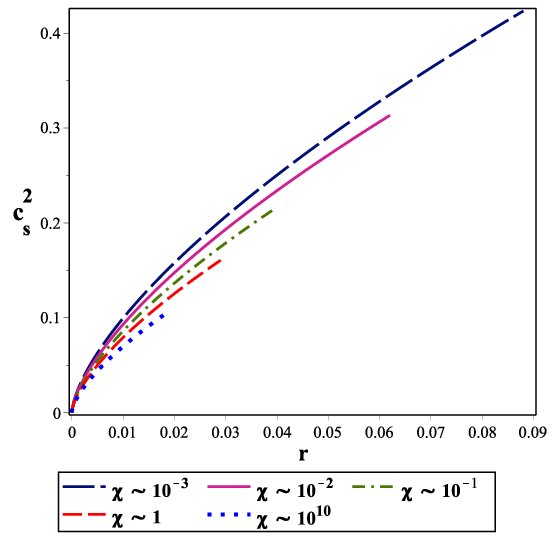}
\includegraphics[scale=0.35]{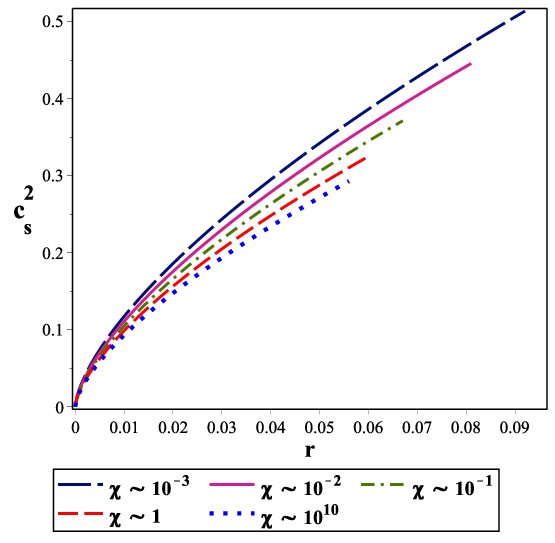}
\end{center}
\caption{\small {Sound speed versus the tensor-to-scalar
ratio in a nonminimal derivative model in which both the nonminimal
coupling and potential are E-model. The left panel corresponds
to $N=60$ and the right panel is for $N=70$.}}
\label{fig6}
\end{figure}

\begin{figure}[]
\begin{center}
\includegraphics[scale=0.57]{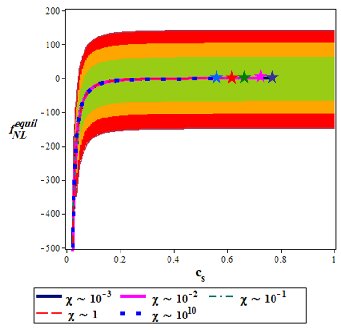}
\includegraphics[scale=0.59]{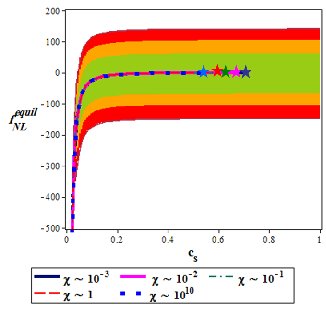}
\end{center}
\caption{\small {Amplitude of the equilateral
configuration of the non-Gaussianity versus the sound speed in a
nonminimal derivative model in which the nonminimal coupling is a
constant and the potential is E-model. The left panel corresponds to $N=60$ and the right panel is for
$N=70$. Note that, here the stars show the maximum values of the
sound speed in each case.}}
\label{fig7}
\end{figure}

\begin{figure}[]
\begin{center}
\includegraphics[scale=0.58]{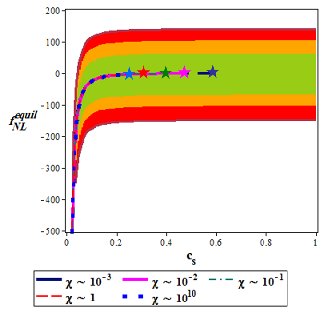}
\includegraphics[scale=0.59]{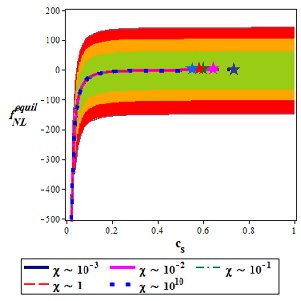}
\end{center}
\caption{\small {Amplitude of the equilateral
configuration of the non-Gaussianity versus the sound speed in a
nonminimal derivative model in which both the nonminimal coupling
and potential are E-model. The left panel corresponds to $N=60$
and the right panel is for $N=70$. The stars show the maximum values of the sound speed in each case.}}
\label{fig8}
\end{figure}

\section{Summary}

The aim of this paper was to study the non-minimal derivative model
in the context of $\alpha$-attractor. In this regard, we have
considered a two-fields action in which both the fields and their
derivatives are nonminimally coupled to the gravity. We have used
the conformal gauge to convert the two-fields model into a one field
which its derivative (and not the field itself) is nonminimally
coupled to the gravity. After obtaining the the background equations
of this NMDC model, we have studied the linear perturbations theory.
By adopting the ADM formalism, we have studied both the scalar and
tensor perturbations in this setup and found some important
perturbation parameters like as the power spectrum, scalar spectral
index and tensor-to-scalar ratio. After that, by introducing the
types of the general functions $\hat{{\cal{F}}}(\psi,\varphi)$ and
$U(\psi,\varphi)$, and using the conformal gauge, the NMDC function
and potential have been obtained. In this paper, two types of NMDC
function have been considered; ${\cal{F}}=const.$ and
${\cal{F}}={\cal{F}}(\phi)$. We have shown that in both cases there is
an attractor \textit{line} in large $N$ and small $\alpha$ limits which
the $r-n_{s}$ trajectories tend to, while in the
the single field $\alpha$-attractor model there is an attractor
\textit{point} in these limits. In fact, the
presence of attractor line instead of attractor point is due to the
nonminimal derivatives coupling that causes the scalar spectral
index of this model to be a functions of two parameters, $\alpha$ and $M$.
In the limit $M\rightarrow \infty$, one recovers the attractor \emph{point} as is usual
in $\alpha$-attractor models. We have numerically studied the linear
perturbation in two cases with constant NMDC coupling and E-model
NMDC coupling (in both cases the potential is considered to be
E-model). With these choices, the power spectrum of the
perturbations in NMDC model can get the observationally viable value
(almost $2.4\times10^{-9}$) in some ranges of $\chi$ and $\alpha$.
Also, in this model $r-n_{s}$ plane lies in the background of
Planck2015 TT, TE, EE+lowP data for all values of $M$ (or $\chi$)
and in some ranges of $\alpha$, for both $N=60$ and $N=70$. We have
obtained some constraints on the free parameter $\alpha$ which lead
to the values of the scalar spectral index and tensor-to-scalar
ratio which are consistent with $68\%$ and $95\%$ CL of the
Planck2015 TT, TE, EE+lowP data for $r-n_{s}$ distribution. By
studying the nonlinear perturbation and three point correlation
functions, we have studied the non-Gaussian feature of the
primordial perturbations. In this regard we have obtained the
amplitude of the equilateral configuration of the non-Gaussianity in
terms of the sound speed. The sound speed is related to the
tensor-to-scalar ratio via the consistency relation. The
tensor-to-scalar ratio is constrained by using the $95\%$ CL of the
Planck2015 TT, TE, EE+lowP data for $r-n_{s}$ distribution. By using
the constraints on the tensor-to-scalar ratio, we have obtained some
constraints on the values of the sound speed in this model. In this regard, we
have shown that it is possible to have small sound speed leading to
the large non-Gaussianity in this NMDC $\alpha$-attractor model.\\

{\bf Acknowledgement}\\
We would like to appreciate the referee for very insightful comments that have
boosted the quality of the paper considerably. This work has been supported financially by Research
Institute for Astronomy and Astrophysics of Maragha (RIAAM) under research project number 1/5237-97.

\end{document}